\documentstyle[12pt]{article}

\begin{document}

\vspace{2cm}
\newcommand{\be}{\begin{equation}}
\newcommand{\ee}{\end{equation}}
\newcommand{\te}{\textstyle}

\centerline{\bf Vibrational Aspects of the  SU(2) Skyrmion}
\vspace{.4cm}
\centerline{F. M. Steffens \footnote{Present address: Department of
Physics and Mathematical Physics, University of Adelaide,
Adelaide, S.A. 5005, Australia} and V.E. Herscovitz}
\date{April 1993}

\vspace{1cm}
\centerline{Instituto de F\'{\i}sica}
\centerline{Universidade Federal do Rio
Grande do Sul}
\centerline{C.P. 15051, 91500, Porto Alegre, RS, Brasil.}
\vspace{.3cm}
\begin{abstract}

We treat the Skyrme model with the breathing mode in a situation
involving two quartic terms. It is seen that there is a new limit
for large $e$ due to the breathing mode not found in the usual rotating
hedgehog.

\end{abstract}

\newpage

\section{Introduction}

Recently many discussions have appeared in the literature [1 -4] about the
stability
of a chiral soliton without the Skyrme term, in particular the
soliton solution to the nonlinear sigma model by means of the so called
breathing mode (BM) \cite{jain,carlson}. In this kind of solution, the radial
variable,
present in the hedgehog solution, is scaled by a dynamical variable so that its
subsequent quantization furnishes an extra energy for the system. The proposal
is
that this energy is responsible
for stability in a mechanism similar to the hydrogen atom. In fact, because
of the BM, one can see in a rough approximation, using the
uncertainty principle, that the classical energy (and radius), even when the
rotations are absent, has a minimum. Although the results are good and
comparable to that of the Skyrme model \cite{skyrme,adk1} (which are obtained
when as well as the nonlinear sigma term, the Skyrme term is also
present) there are some problems that remain, one of them being the
choice of the profile function $F(r)$ (determined in the Skyrme model by the
minimization of the static mass). When the Skyrme term is absent, the static
mass has no minimum and the profile function is quasi-arbitrarily chosen. This
fact can lead to a new kind of instability as one can choose a function that
results in divergent integrals even when the required boundary conditions are
obeyed \cite{kobayashi}, so that the presence or not of the BM
stabilizer term would make no difference.

As the BM has shown its importance in the non\-li\-ne\-ar si\-gma model
through the existence of a special stable soliton configuration, one can argue
what is its relevance in the Skyrme model (for example, how accurately one
can describe the nucleon and delta resonances). Although there are already
some treatments of this subject in the literature \cite{hajduk,asano}, mainly
about the stability \cite{sawada} of the Skyrme model with a vibrational
solution, here we intend to explore the relationship between the Skyrme model
and the nonlinear sigma model, more specifically how one can get the results
of the nonlinear sigma model with BM as was done, for example, by
Carlson \cite{carlson}, from the resolution of the Schr\"odinger equation of
the Skyrme model with BM. In other words, we depart from the
Skyrme model, which has a well defined profile function, perform the
quantization of the rotational and vibrational degrees of freedom and take
a suitable limit on the subsequent Schr\"odinger equation. As a
result, we get  the vibrating nonlinear sigma model with finite integrals. This
kind
of limit is a property only of the vibrating skyrmions. The same limit exists
for all static properties.
This is the main idea of the present contribution.

\section{The breathing mode}

The SU(2) Skyrme model in a version that includes the two independent
fourth order terms (Skyrme and symmetric quartic terms (SQT)) and in the
chiral symmetry limit is described by the following Lagrangian
density ($\hbar=c=1$):

\begin{equation}
{\cal L}_{sk}= -\frac{F_\pi^2}{16}Tr(L_\mu L^\mu) +
\frac{1}{32e^2}Tr([L_\mu ,L_\nu]^2) +
\frac{\gamma}{8e^2}\{[Tr(L_{\mu}L^{\mu})^2
]\},
\label{a2}
\end{equation}
with $L_{\mu}=U^{\dagger}\partial_{\mu}U$, $F_{\pi}$=186 MeV the pion decay
constant, $e$ and $\gamma$ parameters to be adjusted and
$U=e^{i\vec\tau\cdot\vec\pi(\vec x,t)}$ the pion field. Skyrme
introduced soliton solutions in his model through the field
configuration called hedgehog.
In this ansatz, the pion field has a radial direction and depends on
a function $F(r)$ called the chiral angle. This function is determined
by the minimization of the static mass associated with the hedgehog
solution. Well defined states of spin and isospin are obtained
through a rotation of the classical solution.
The BM can be taken into account by means of a scaling on the
radial variable $r$:
\be
U=A(t)e^{i\vec\tau\cdot\hat{r}F(r,R(t))}A^{\dagger}(t),
\label{a5}
\ee
where $R(t)$ is the dynamical variable that characterizes the soliton breathing
mode and $A(t) \in$ SU(2) is the rotation matrix. The use of the solution
(\ref{a5}) in the Lagrangian density (\ref{a2})
and an integration over all space leads to the following Lagrangian:
\be
L=R^3 b\dot a_{\mu}\dot a^{\mu} + R(k - \gamma m)\dot a_{\mu}\dot a^{\mu} - Rc
-\frac{h - \gamma l}{R} + \dot R^2 (f - \gamma n) + R\dot R^2 d,
\label{a6}
\ee
with $a_{\mu}$ the four variables that parametrize the rotation matrix $A(t)$
and the constants are listed in appendix. We stress that
the integrals $c$, $b$ and $d$ are the contributions from the nonlinear
sigma term to the static mass, to the rotational term and to the BM,
respectively. In the same order, $h$, $k$ and $f$ are the
contributions from the Skyrme term and $l$, $m$ and $n$ the contributions
from the symmetric quartic term.

One can pass to the Hamiltonian formulation through a Legendre transformation,
whose result is:
\be
H=\frac{\pi^2}{4R(R^2 b + k - \gamma m)} + \frac{p_{R}^{2}}{4(Rd + f - \gamma
n)}
+ Rc + \frac{h - \gamma l}{R},
\label{a7}
\ee
with $\pi^2=\pi_{\mu}\pi^{\mu}$ the square of the momentum conjugate to
$a_{\mu}$
and $p_R$ the momentum conjugate to $R$.
Now, we will write the Hamiltonian (\ref{a7}) through a set of reduced
variables
obtained when we make use of the constraint on the rotational variables
($a_{\mu}a^{\mu}=1$). We remind the reader that if we first
make the reduction of variables at the classical level and perform
the quantization after,  or first perform the quantization and afterwards
restrict
adequately the vectors of the Hilbert space, we obtain the same result
\cite{zahed,eu}.
Another way is to construct a Hamiltonian compatible with the constraint by
means
of Dirac's procedure for constrained systems \cite{dirac}. In this case
there appears
a contribution to the zero point energy inversely proportional to the moment of
inertia \cite{eu,outros}. This contribution will not be treated in the present
work.

Using the reduced coordinates $q=(a^i , R)$ and $p=(\pi_i ,p_R)$, $i$=1,2,3,
we write the Hamiltonian as:
\be
H=\frac{1}{2}p_a g^{ab}p_b + cq^4 + \frac{h - \gamma l}{q^4},
\label{a8}
\ee
where:
\be
g^{ab}=\left[
 \begin{array}{cc}
 \frac{\te{g^{ij}}}{\te{R(R^2 b + k - \gamma m)}} & 0 \\
 0 & \frac{\te{1}}{\te{2(Rd + f - \gamma n)}}
 \end{array}
 \right]
\label{a9}
\ee
and
\be
g^{ij}=\frac{1}{2}(\delta^{ij}-q^i q^j).
\label{s32}
\ee

We write the quantum Hamiltonian via the DeWitt \cite{dewitt} prescription:
\begin{eqnarray}
H\psi &=& -\frac{1}{2g^{1/2}}\partial_a (g^{1/2}g^{ab}\partial_b \psi) +
\left(q^4 c\psi + \frac{h - \gamma l}{q^4}\psi\right) \nonumber \\*
      &=& -\frac{1}{2g^{1/2}}\partial_i (g^{1/2}g^{ij}\partial_j \psi)
          -\frac{1}{2g^{1/2}}\partial_4 (g^{1/2}g^{44}\partial_4 \psi)
          + Rc\psi + \frac{h - \gamma l}{R}\psi
\label{a10}
\end{eqnarray}
or
\begin{eqnarray}
      &\space& -\frac{1}{4d}\left\{\frac{\partial^2}{\partial R^2} +
           \frac{4R^4 + (5(f - \gamma n)/d + (k - \gamma m)/b)R^2 +
           2(f - \gamma n)(k - \gamma m)/bd}{R(R^2 + (f - \gamma n)/d)(R^2 + (k
- \gamma m)/b)}
           \frac{\partial}{\partial R}\right\}\psi \nonumber \\*
      &\space& + \frac{R^2 + (f - \gamma n)/d}{R}\left(\frac{j(j+1)}{R(bR^2 +
           k - \gamma m)}
           + Rc + \frac{h - \gamma l}{R}\right)\psi = \frac{R^2 + (f - \gamma
n)/d}{R}E\psi ,
\label{a11}
\end{eqnarray}
where $g$ is the determinant of the metric tensor.
We make a transformation on the wave function $\psi$ as indicated in
\cite{asano},
\be
\psi=\left\{\frac{2(dR^2 + f - \gamma n)}{R}\right\}^{1/4}\left\{2R(bR^2 +
k - \gamma m)\right\}^{-3/4}
\phi .
\label{a12}
\ee
and eliminate the first derivative from the equation.
So:
\begin{eqnarray}
&\space&-\frac{1}{4d}\frac{\partial^2}{\partial R^2}\phi \nonumber \\*
&\space&+\frac{8R^6 + (18(f - \gamma n)/d + 10(k - \gamma m)/b)R^4}
        {16d\left((f - \gamma n)/d + R^2\right)^2 \left((k - \gamma m)/b +
R^2\right)^2}\phi \nonumber \\*
&\space&+\frac{(15(f - \gamma n)^2 /d^2 + 18(f - \gamma n)(k - \gamma
m)/db)R^2}
        {16d\left((f - \gamma n)/d + R^2\right)^2 \left((k - \gamma m)/b +
R^2\right)^2}\phi \nonumber \\*
&\space&-\frac{(k - \gamma m)^2 R^2 /b^2 +18(k - \gamma m)(f - \gamma n)^2
/bd^2
        - 6(f - \gamma n)(k - \gamma m)^2
        /db^2}{16d\left((f - \gamma n)/d + R^2\right)^2 \left((k - \gamma m)/b
+ R^2\right)^2}\phi
\nonumber \\*
&\space& + \frac{(R^2 + (f - \gamma n)/d)}{R}\left(\frac{j(j+1)}{R(bR^2 + k -
\gamma m)}
           + cR + \frac{h - \gamma l}{R}\right)\phi \nonumber \\*
&\space&= \frac{R^2 + (f - \gamma n)/d}{R}E\phi .
\label{a13}
\end{eqnarray}
In order to obtain a solution to this equation, we observe that the behavior
of $\phi$ near the origin is:
\be
\phi(R\rightarrow 0)\sim R^{\frac{\te 1}{\te 2} + \left\{\frac{\te 1}{\te 4}
+ {\te 4(f - \gamma n)}\left(\frac{\te j(j+1)}{\te k - \gamma m} +
\te{h - \gamma l}\right)\right\}^{\te 1/2}}
\label{a14}
\ee
and we require that $\phi$ goes to zero at infinity.

\section{The limit of large $e$}

Usually in the Skyrme model, the two free parameters of the theory are
chosen to reproduce the nucleon and delta masses. As a result, the value of $e$
that reproduces such masses does not correspond to the $e$ value that minimizes
them. This is not so important in the context of the value of $e$ to be used,
because there is no necessity to use the $e$ that minimizes the mass. But it is
important in the sense that it shows that for $e$ going to zero or $e$ going to
infinity, the mass diverges (see figure \ref{fig1}). Now, the question is what
happens with the mass
in that limit when along with the rotations,  the BM is also present?
The answer is that for small $e$ the mass still diverges, but for
sufficiently large $e$, the mass becomes a constant in
$e$ and all works as if only the nonlinear sigma term were present, with a
profile function $F$ that minimizes the static mass of the Skyrmion.

The results
for the nucleon and delta masses and their excitations are the same as those
obtained some time
ago by Carlson \cite{carlson}. In that work, Carlson pointed out that it is
possible
to obtain stable soliton configurations from the nonlinear sigma model when the
Skyrme
term is used as a constraint in the action of the system. This was shown using
scale invariant arguments in the functional integral
for the field configurations $U$, dividing the configurations into equivalence
classes and extending the functional integral to the equivalence classes of
configurations. The resulting functional integral depends of a dilatated field
and involves a restriction on a particular function $G(U)$ of $U$. This
function must preserve the chiral limit and must not be invariant by
dilatations as, in this case, it would not be possible to choose a field
configuration that was representative of its class. Carlson chose for $G(U)$
the Skyrme term. With this choice, it is possible to solve the Euler-Lagrange
equation without the explicit presence of the Skyrme term in the action. So,
Carlson obtains a hamiltonian of the nonlinear sigma model with integrals
dependent of the chiral angle from the Skyrme term.

Now, we will call $y$ the dimensionless variable that scales
the radial variable in the nonlinear sigma model. It is related to $R$
by $R=ey$. If one makes this replacement in the Schr\"odinger equation
(\ref{a11}) is possible to verify that for $e\rightarrow\infty$, it
reduces to:
\be
 -\frac{1}{4yd}\left\{\frac{\partial^2}{\partial y^2} +
 \frac{4}{y}\frac{\partial}{\partial y}\right\}\psi +
 \frac{j(j+1)}{by^3}\psi + cy\psi = E\psi ,
\label{a15}
\ee
which is $e$ independent. As the SQT and the Skyrme term have the same $e$
dependence, the equation has the same functional form when $\gamma$ is zero
or not in the limit of large $e$, with the only difference being the function
$F$ that minimizes the classical mass. The expression (\ref{a15})
is the Schr\"odinger equation for the BM when only the
nonlinear sigma term is present. In fact, the numerical resolution of equation
(\ref{a11}) shown in figure (\ref{fig1}), tells us that for $e\geq_{~}$ 80 the
mass is already that calculated by Carlson. The other static properties have
the same kind of limit. As an example, we
write down the expression for the axial coupling $g_a$. It is easy to verify
that $g_a$, given in the vibrating Skyrme model by

\begin{eqnarray}
g_a&=&\frac{1}{e^2}\frac{\pi}{3}\frac{\int R^2\mid\psi\mid^2 \sqrt{g}dR}{\int
\mid\psi\mid^2 \sqrt{g}dR}
\int_{0}^{\infty}d\rho \rho^2\left(F' + \frac{sin(2F)}{\rho}\right) \nonumber
\\*
&\space&  + \frac{8\pi}{3e^2}\int_{0}^{\infty}d\rho sin^2 FF'\nonumber \\*
&\space&  + \frac{4\pi}{3e^2}\left(\int_{0}^{\infty}d\rho \rho sin(2F)(F')^2 +
\int_{0}^{\infty}d\rho sin^2 F\frac{sen(2F)}{\rho}\right),
\label{a16}
\end{eqnarray}
is  reduced, in the limit of large $e$, to:
\be
g_a = \frac{\pi}{3}\frac{\int y^2\mid\psi\mid^2 \sqrt{g}dy}{\int
\mid\psi\mid^2 \sqrt{g}dy}
\int_{0}^{\infty}d\rho \rho^2\left(F' + \frac{sin(2F)}{\rho}\right),
\label{a17}
\ee
which is the corresponding expression in the nonlinear sigma model. The same
holds for all other static properties.

\section{Numerical results}

Figure
(\ref{fig1}) shows the mass curve for the case of the rotating hedgehog with
the
Skyrme term (case I), Skyrme term plus BM (case II) and Skyrme and SQT terms
plus
BM (case III). The
value used for the pion coupling constant is $F_{\pi}=186$ MeV and $\gamma$ is
0.11
which is the maximum $\gamma$ value that mimimizes
the  classical mass. This value agrees with those found in the literature
\cite{lacombe}. The discrepancy between cases II and III is due to the presence
or absence of the SQT. As seen in equation (\ref{a13}), the SQT contributes
negatively
to the energy so that for low $e$, the mass calculated when the SQT is present
is lower than that calculated when only the Skyrme term is present. In the high
$e$ region, the only difference is in the constants $b$, $c$ and $d$ of
equation
(\ref{a15}) which are calculated with different profile functions. For case II
they
are 125.01/$F_{\pi}e^3$, 18.23$F_{\pi}$/$e$ and
160.79/$F_{\pi}e^3$ and for case III 52.57/$F_{\pi}e^3$, 14.72$F_{\pi}$/$e$ and
72.87$/F_{\pi}e^3$ respectively (see table in the appendix). The asymptotic
mass
value for case II is 1101 MeV
and for case III 1152 MeV compared with an experimental value given by 939 MeV.

As the energy is not absolutely defined in our treatment, we decided to search
for mass differences (fig. \ref{fig2}). Initially we have two free parameters
in the model. One of them, as stated before,
was chosen to
fit the pion decay constant. The other, in principle, could be chosen to fit
the mass diference between the states with $j$=1/2 and $j$=3/2. The problem is
that
for small $e$ the mass difference is too small and for large $e$ it is
independent of $e$ so that in this region we have only one free parameter. As
we choose
to fit $F_{\pi}$, the biggest mass difference for case II is 285 MeV and for
case III is 306 MeV. The experimental value is
293 MeV. The differences between cases II and III arise because the centrifugal
term
in III is bigger than II implying that the mass difference is also bigger.
In figure (\ref{fig2}) we also plot the mass difference for the
rotating hedgehog with the Skyrme term. As is well known, it behaves like
$e^3$.
Finally, to illustrate the behaviour of the other static properties,
the isoscalar radius curve is plotted in figure (\ref{fig3}). One can see that
for large $e$, the
isoscalar radius for the rotating hedghog goes to zero, as expected, because
in that region the effect of the Skyrme term is weak and the soliton collapses.
On the other side, when the BM is present, it is verified that for large
$e$ the isoscalar radius goes to a constant value. For case II, the mean
charge radius is 0.56 fm and for case III is 0.58 fm. Another example is
the isoscalar mean magnetic radius. For case II, it is 0.96 fm and for case
III is 0.92 fm.

\section{Conclusions}

We have studied the Skyrme model in a solution for
the pion field that includes both rotational and vibrational
degrees of freedom. It was found that the presence of the breathing
mode in the solution allows one to take the limit $e\rightarrow\infty$ -
a situation not found in the usual rotating hedgehog. As a consequence,
we recovered the earlier results of Carlson \cite{carlson}. We did this
study in two situations: one involving only the Skyrme term and
the other involving the Skyrme term plus the symmetric quartic term.
The limit for large $e$ is the same for the two cases with the only
difference being the function $F$ that minimizes the classical mass.
Some static properties were presented and it was noticed that the masses
of the nucleon and delta states as well
as the mass difference between them are larger when the symmetric quartic term
is present. This is due to centrifugal
effects.
We also think that the method presented here provides some justification
for the use of a particular chiral angle in the model proposed by
P. Jain et al. \cite{jain}.
\newline
\newline
\newline

One of us (F.M.S.) acknowledges helpful discussions with M. Betz and G. L.
Thomas.
This work was supported by the Brazilian agencies CNPq, FAPERGS and
FINEP.

\appendix
\renewcommand{\thesection}{}
\section{Appendix: Integrals containing the chiral angle}
\renewcommand{\thesection}{\Alph{section}}
\setcounter{equation}{0}

Let $F(\rho)$ be the chiral angle,
$s=sinF$, $c=cosF$, $F_{\pi}$ the pion decay constant
($\pi\rightarrow$ $\mu$ + $\nu$) and $e$ the Skyrme dimensionless parameter.
So, the explicit form for the constants appearing in the Lagrangean is  given
by:

\be
b=\frac{4\pi}{3e^3 F_{\pi}}\int_{0}^{\infty}s^2 \rho^2d\rho,
\ee
\be
c=\frac{\pi F_{\pi}}{2e}\int_{0}^{\infty}\left(\left(\frac{\partial F}
{\partial\rho}\right)^2 + \frac{2s^2}{\rho^2}\right)\rho^2 d\rho,
\ee
\be
d=\frac{\pi}{2e^3 F_{\pi}}\int_{0}^{\infty}\left(\frac{\partial F}
{\partial\rho}\right)^2 \rho^4 d\rho,
\ee
\be
f=\frac{4\pi}{e^3 F_{\pi}}\int_{0}^{\infty}s^2 \left(\frac{\partial F}
{\partial \rho}\right)^2 \rho^2 d\rho,
\ee
\be
g=\frac{16\pi}{3e^3 F_{\pi}}\int_{0}^{\infty}s^2\left(\left(\frac{\partial
F}{\partial\rho}
\right)^2 + \frac{s^2}{\rho^2}\right)\rho^2 d\rho,
\ee
\be
h=\frac{2\pi F_{\pi}}{e}\int_{0}^{\infty}s^2\left (2\left(\frac{\partial
F}{\partial\rho}
\right)^2 + \frac{s^2}{\rho^2}\right) d\rho,
\ee
\be
l=\frac{2\pi F_{\pi}}{e}\int_{0}^{\infty} \left(\left(\frac{\partial
F}{\partial\rho}
\right)^2 + 2\frac{s^2}{\rho^2}\right)^2 \rho^2 d\rho,
\ee
\be
m=\frac{32\pi}{3e^3 F_{\pi}}\int_{0}^{\infty}s^2 \left(\left(\frac{\partial F}
{\partial\rho}\right)^2 + 2\frac{s^2}{\rho^2}\right) \rho^2 d\rho,
\ee
\be
n=\frac{4\pi}{e^3 F_{\pi}}\int_{0}^{\infty} \left(\frac{\partial
F}{\partial\rho}
\right)^2 \left(\left(\frac{\partial F}{\partial\rho}\right)^2
+ 2\frac{s^2}{\rho^2}\right) \rho^4 d\rho,
\ee

The coefficients $c$, $h$ and $l$ are the contributions for the soliton
static mass from the nonlinear sigma term, the Skyrme term and SQT
respectively.
In the same way, the coefficients $b$, $g$ and $m$ are the contributions for
the rotational energy and  $d$, $f$ and $n$ the contributions for the
vibrational energy.

\begin{table}
\begin{center}
 \begin{tabular}{|c|c|c|}
 \hline
 \sl Integrals & \sl $\gamma=0$ & \sl $\gamma=0.11$ \\
 \hline
 b & 125.0145 & 52.573 \\
 c & 18.23082 & 14.716 \\
 d & 160.7893 & 72.874 \\
 f & 43.91    & 39.994 \\
 g & 88.52515 & 75.37  \\
 h & 18.23    & 22.965 \\
 l &    -     & 74.988 \\
 m &    -     & 194.834 \\
 n &    -     & 193.077 \\
 \hline
 \end{tabular}
\end{center}
\caption{Numerical values for the integrals containing the chiral angle  when
only the
Skyrme term is present ($\gamma$=0) and when the SQT ($\gamma$=0.11) is added.
The parameters
 $e$ and $F_{\pi}$ are not included.}
\label{figa1}
\end{table}

\addcontentsline{toc}{chapter}{\protect\numberline{}{References}}

\newpage

\begin{figure}
\vspace{5cm}
\special{fig1.ps}
\caption{Energy curve for the rotating hedgehog with and without BM
in function of the Skyrme parameter $e$. In cases II (dashed line) and III
(continuous line), where the BM
is present, the energy decreases with $e$ until a region where it becomes $e$
independent. In case I (dotted line), the energy diverges as $e$ tends to zero
or
infinity.}
\label{fig1}
\end{figure}

\begin{figure}[h]
\vspace{8cm}
\special{fig2.ps}
\caption{Mass difference curve between the states $j$=1/2 and $j$=3/2 for the
rotating hedgehog with and without BM in function of the Skyrme parameter $e$.
In cases II (dashed line) and III (continuous line), the mass difference
increases
until a region where it
becomes $e$ independent. In case I (dotted line), the mass difference diverges
as $e$ goes to
the infinity.}
\label{fig2}
\end{figure}

\begin{figure}[h]
\vspace{8cm}
\special{fig3.ps}
\caption{Isoscalar radius curve for the rotating hedgehog with and without
breathing mode in function of the Skyrme parameter $e$. The notation is the
same
as in figure 1. In the case that the
breathing mode is present, the isoscalar radius decreases with $e$ until it
reaches a minimum and tends to a constant. In the other case, the radius goes
continuously to zero as the strength of the Skyrme term becomes weaker.}
\label{fig3}
\end{figure}


\begin{thebibliography}{40}
\def\bi{\bibitem}

\bi{jain} P. Jain, J. Schechter and R. Sorkin, {\it Phys. Rev.} {\bf D39}
(1989),
998; {\bf 41} (1990), 3855;
R. K. Bhaduri, A. Susuki, A. H. Abdalla and M. A. Preston, {\it Phys. Rev.}
{\bf D41} (1990), 959;
N. M. Chepilko, K. Fujii and A. P. Kobushkin, {\it Phys. Rev.} {\bf D43}
(1991), 2391;
P. Jain, {\it Phys. Rev.} {\bf D41} (1990), 3527;
B. S. Balakrishna, V. Sanyuk, J. Schechter and A. Subbaraman,
{\it Phys. Rev.} {\bf 45} (1992), 344.

\bi{carlson} J. W. Carlson, {\it Nucl. Phys.} {\bf B253} (1985), 149; {\bf
B277}
(1986), 253.
\bi{mignaco} J. A. Mignaco and S. Wulck, {\it Phys. Rev. Lett.} {\bf 62}
(1989), 1449;
{\it J. Phys.} {\bf G18} (1992), 1309.
\bi{okazaki} T. Okazaki, K. Fujii and N. Ogawa, {\it Int. J. of Mod. Phys.}
{\bf A27}
(1992), 6763.
\bi{skyrme} T.H.R. Skyrme, {\it Proc.Roy.Soc.} {\bf A260}, 127;{\bf A262},
(1961) 237;
{\it Nucl.Phys.} {\bf 31} (1962), 550; {\it ibid.} 556.

\bi{adk1} G.S. Adkins, C.R. Nappi and E. Witten, Nucl.Phys. {\bf
B228}, 552 (1983); E. Witten, Nucl.Phys. {\bf B223}, 422 (1983);
G.S. Adkins, in {\it Chiral Solitons}, K-F-Liu ed., World
Scientific, p.99 (1987).

\bi{kobayashi} A. Kobayashi, H. Otsu and S. Sawada, {\it Phys. Rev.} {\bf D42}
(1990), 1868.

\bi{hajduk} C. Hajduk and B. Schwesinger, {\it Phys. Lett.} {\bf 140B} (1984),
172.
\bi{asano} H. Asano, H. Kanada and H. So, {\it Phys. Rev.} {\bf D44} (1991),
277.

\bi{sawada} S. Sawada and K. Yang, {\it Phys. Rev.} {\bf D44} (1991), 1578.







\bi{zahed} I. Zahed and G. E. Brown, Phys. Rep. {\bf 142} (1986), 1.
\bi{eu} F. M. Steffens, M. Sc. dissertation, UFRGS - Univ. Fed. do Rio Grande
do Sul (Brazil), 1992.
\bi{dirac} P.A.M Dirac, {\it Lectures on Quantum Mechanics}, Ye\-shi\-va
University, New York (1964).

\bi{outros} N. Ogawa, K. Fujii and A. Kobushkin, {\it Prog. Theor. Phys.} {\bf
83},
(1990) 894; {\bf 85} (1991), 1189;  H. Vershelde and H. Verbek, {\bf Nucl.
Phys.}
{\bf A500} (1989), 573.


\bi{dewitt} B.S. DeWitt, {\it Phys. Rev.} {\bf 85} (1952), 653.

\bi{lacombe} J. F. Donoghue, E. Golowich and B. R. Holstein, {\it Phys. Rev.
Lett.}
{\bf 53} (1984), 747;
T. N. Pham and T. N. Truong, {\it Phys. Rev.} {\bf D31} (1985), 3027;
M. Lacombe, B. Loiseau, R. Vinh Mau and W. N. Cottingham, {\it Phys. Lett.}
{\bf 161B} (1985), 31.

\end{thebibliography}
\end{document}